\documentstyle[12pt]{article}
\newcommand{\be}{\begin{eqnarray}}
\newcommand{\ee}{\end{eqnarray}}

\catcode`\@=11
\def\lsim{\mathrel{\mathpalette\@versim<}}
\def\gsim{\mathrel{\mathpalette\@versim>}}
\def\@versim#1#2{\vcenter{\offinterlineskip
\ialign{$\m@th#1\hfil##\hfil$\crcr#2\crcr\sim\crcr } }}
\catcode`\@=12

\begin{document}

\pagestyle{empty}

\noindent
\hspace*{10.7cm} \vspace{-3mm}  CERN-TH/2000-084\\
\hspace*{10.7cm} \vspace{-3mm}  FERMILAB-Pub-00/013-T\\
\hspace*{10.7cm} \vspace{-3mm}  HIP-1999-76/TH\\
\hspace*{10.7cm} \vspace{-3mm}  KUNS-1650\\


\begin{center}
{\Large\bf RG-invariant Sum Rule 
in a Generalization of\\
 Anomaly Mediated SUSY Breaking Models} 
\end{center} 

\vspace{1cm}
\begin{center}
{\sc Marcela Carena}$\ ^{(a,b)}$, 
{\sc Katri Huitu}$\ ^{(c)}$ and
{\sc Tatsuo Kobayashi} $\ ^{(d)}$
\end{center}
\begin{center}
{\em $\ ^{(a)}$ 
Theory Division, CERN, CH-1211 Geneva 23, Switzerland} \vspace{-2mm}\\
{\em $\ ^{(b)}$ 
Fermi National Accelerator Laboratory, Batavia, IL 60510, 
USA} \vspace{-2mm}\\
{\em $\ ^{(c)}$ 
Helsinki Institute of Physics, 
FIN-00014 University of Helsinki, Finland} \vspace{-2mm}\\
{\em $\ ^{(d)}$ 
Department of Physics, Kyoto University, Kyoto 606-8502, 
Japan} \vspace{-2mm}\\
\end{center}

\vspace{1cm}
\begin{center}
{\sc\large Abstract}
\end{center}

\noindent
We study a generalization of  anomaly-mediated supersymmetry 
breaking (AMSB) scenarios, under the assumption that the effects of the
high-scale theory do not completely decouple and D-term type
contributions can be therefore present. 
We investigate the effect of such possible D-term 
 additional contributions to soft scalar masses by 
requiring that, for non-vanishing, renormalizable Yukawa couplings
$Y^{ijk} $, the sum of squared soft supersymmetry breaking mass parameters, 
$M^2_{ijk} \equiv m_i^2+m_j^2+m_k^2$, is RG-invariant, in the sense
that it becomes independent of the specific ultraviolet boundary
conditions as it occurs in the AMSB models.
These type of models can avoid the problem of  tachyonic solutions for the
slepton mass spectrum present  in AMSB scenarios.
We implement the  electroweak symmetry 
breaking condition and explore the sparticle spectrum
associated with this framework. To show the possible diversity of the
sparticle spectrum, we consider two examples, one in which the D-terms
induce a common  soft supersymmetry breaking mass term for
all sfermion masses,  and another one in which a light stop
can be present in the spectrum.

\newpage
\pagestyle{plain}

\section{Introduction}

Supersymmetry(SUSY)  provides a well-motivated
extension of the  Standard Model (SM) with an elegant solution to the 
so-called naturalness problem associated to the SM Higgs sector.
In low energy supersymmtric models the electroweak
scale is naturally of the order of the soft SUSY breaking parameters
of the theory.
Much work has been done in the search for an appropriate mechanism 
for SUSY breaking, but, at present, it remains unknown. 
Two types of SUSY breaking mediation mechanisms, 
supergravity-mediated \cite{nilles} and 
gauge-mediated \cite{GR}, have been studied extensively.
Recently, another type of mediation mechanism, i.e. Anomaly--Mediated
Supersymmetry Breaking (AMSB), 
has become under scrutiny \cite{randall,giudice,pomarol}.\footnote{See 
also Ref.\cite{chacko,bagger}. In Ref.\cite{bagger} anomaly mediation 
has been discussed within the framework of supergravity.}
Each of these mechanisms has unique aspects which differ from each other.
One of the important aspects of AMSB is that  
the soft SUSY breaking terms are 
Renormalization Group (RG) invariant,  in the sense
that they become independent of the specific ultraviolet boundary
conditions.
In fact, the magnitudes of 
the  soft SUSY breaking terms at any scale are obtained 
in terms of values of the related gauge and Yukawa couplings at that
scale, hence, they are determined as a function of the measured values
of those couplings at the weak scale.
Thus, the AMSB scenario yields definite
phenomenological predictions. However,
within the framework of the minimal supersymmetric standard model (MSSM), 
one of the predictions is quite problematic, since it implies the
existence of negative values for the 
slepton squared masses.
The simplest way to solve  this problem is to add a universal contribution 
$m_0$ to all soft  SUSY breaking scalar
masses \cite{randall,anom-ph}. There is, 
however, no dynamical explanation for the origin of this term and
there is no obvious reason why extra contributions should not appear as well 
for  the other soft SUSY 
breaking terms: gaugino masses and 
the trilinear Yukawa couplings
 $A_f$ which couple Higgs and scalar fermion fields.
Moreover, the addition of the universal value $m_0$ to all soft scalar 
masses, although provides a solution to the tachyonic spectra, it also
violates the RG-invariance, which is one of the most 
attractive aspects in the AMSB scenario.

Independently of recent works on anomaly-mediated SUSY breaking, 
RG-invariant relations of soft SUSY breaking terms 
have been studied in the literature\cite{rg-inv,kubo}.
The relation to the case of anomaly mediated SUSY breaking  has been 
clarified in Ref.\cite{jj}.
In Ref.\cite{kubo}  the RG-invariant sum rule of soft scalar masses 
has been discussed and its importance has been emphasized. That is, 
the sum of three scalar masses squared, $M^2_{i,j,k} \equiv 
m_i^2+m_j^2+m_k^2$, corresponding to chiral fields for which 
the Yukawa couplings $Y^{ijk} \neq 0$ are allowed, 
e.g. $(i,j,k)=(Q,u,H_2), (Q,d,H_1)$ and $(L,e,H_1)$ \footnote{Q=(u,d) and 
L=($\nu_e$, e) are the SU(2) left handed superfield doublets; 
$H_1$, $H_2$ are the  two Higgs doublets and $u$, $d$, $e$ are 
SU(2) right-handed superfield singlets.},  
is more important to RG-invariance 
than any one of the scalar mass terms independently. This is the case
since such a sum appears in 
 the $\beta$-functions of the Yukawa couplings and soft SUSY breaking
 masses themselves.
Hence, one could allow for additional contributions
to   each of the soft scalar masses, as long as 
the sum itself is not  affected.
For example, this situation can be realized by additional D-term 
contributions, which are  proportional to a charge $q_i$ of the 
field under a broken symmetry \cite{Dterm1,Dterm2}
and such that in the allowed Yukawa couplings 
the charge must be conserved, $q_i+q_j+q_k=0$.
Then, in the sum of the soft SUSY breaking parameters, $M^2_{i,j,k}$, the
D-term contributions cancel each other.

In the present work  we shall investigate the possibility of having a
similar behaviour as the one explained above, assuming a
generalization  of  anomaly-mediated SUSY breaking models with residual,
non-decoupling  effects from extra U(1)'s at a high energy scale. 
In \cite{pomarol} it was concluded that all effects coming from a high
scale theory decouple in pure anomaly mediation in the absence of
light singlets. The authors of ref. \cite{pomarol} explored also
extensions of the AMSB scenarios in which non-decoupling effects
survive at low energies allowing, for example,  for genuine D-term
contributions.
In a generic framework, 
D-term contributions have been proposed
as a  solution to the tachyonic slepton mass problem both in Refs.
\cite{pomarol,katz}. 
Here we  shall analyse
the  features of the particle  spectrum depending on the
specific charge assignments of the  
additional D-term  contributions  to  each of
the soft scalar mass parameters. We shall then 
explore  the regions of the parameter space
for which 
no tachyonic slepton masses appear.
A novel point of the present work is to study the phenomenological aspects 
of these theories, 
making use of the RG-invariant sum of the soft SUSY breaking scalar 
squared masses.
In this way, we preserve
one of the most appealing features of the  anomaly-mediated SUSY
breaking scenarios, namely its high predictivity with induced soft masses 
which are independent of flavour physics, and we  cure the main problem
associated to them, namely the tachyonic solutions for the slepton 
sector.\footnote{There has been 
other attempts of constructing generalizations of AMSB models, 
curing the problem of  tachyonic solutions in the slepton sector and still 
assuring a scale invariance of the solutions, like the anti-gauge mediated 
model of ref. \cite{pomarol} or a complementary proposal in ref. \cite{katz}.} 

The paper is organized as follows.
In Section 2 we define the framework:
we assume additional D-term contributions to soft scalar masses $m^2_i$, 
requiring that the sum $M^2_{i,j,k} \equiv m_i^2+m_j^2+m_k^2$ remains 
RG-invariant, a property shared by AMSB scenarios.
We  implement the  electroweak  symmetry breaking condition and
discuss generic class of models 
with suitable  D-term contributions to investigate the properties 
of the mass spectrum.
In Section 3 we investigate specific models  in detail, imposing the 
present experimental bounds on supersymmetric particle masses. We present
 two examples with fixed charge assignments and study their mass spectra.
Section 4 is devoted to our conclusions.

\section{RG-invariant sum rule and D-term contributions}
\subsection{Sum rule}

In the anomaly-mediated SUSY breaking scenario
the soft SUSY breaking parameters, 
i.e. the gaugino masses $M_\alpha$, the soft scalar masses $m_i$ 
and the $A$-parameters are given by, \cite{randall,giudice,pomarol} 
\begin{eqnarray}
M_\alpha &=& {\beta_{g_\alpha} \over g_\alpha}m_{X}, 
\label{gaugino-M} \\
m_i^2 &=& -{1 \over 4}
\left(\sum_\alpha {\partial \gamma_i \over \partial g_\alpha} 
\beta_{g_\alpha}+\sum_{Y^{ijk}} 
{\partial \gamma_i \over \partial Y^{ijk}} 
\beta_{Y^{ijk}}\right)m_{X}^2,
\label{scalar-m}\\
A_{ijk} &=& -{\beta_{Y^{ijk}} \over Y^{ijk}}m_{X},
\label{A-term}
\end{eqnarray}
where $g_\alpha$, with $\alpha=1,2,3$,  and $Y^{ijk}$, with 
 $(i,j,k)=(Q,u,H_2)$, $(Q,d,H_1)$, $(L,e,H_1)$,  are the gauge couplings 
and the Yukawa couplings, and $\beta_{g_\alpha}$ 
and $\beta_{Y^{ijk}}$ are their $\beta$-functions, 
respectively.
Here $\gamma_i$ are the anomalous dimensions of the chiral superfields.
For explicit calculations, it is convenient to define 
$m_F \equiv m_{X}/(16\pi^2)$, because $\beta$-functions 
include the loop-factor $16 \pi^2$.

Eqs.(\ref{gaugino-M})-(\ref{A-term}) are RG-invariant, that is, they
 are valid at any scale.
Thus, the soft SUSY breaking terms are expressed as a function
of gauge and Yukawa couplings at a given scale   
times  the overall magnitude $m_F$.
The consequent high predictability in the model leads to a problem
in the MSSM, since the  $\beta$-function coefficients for the weak
gauge couplings,
$b_1 = \frac{33}{5}$,  
$b_2 = 1$ and $b_3 = -3$
 render the squared soft SUSY breaking parameters for the
sleptons negative, yielding  tachyonic slepton masses.
Here we shall discuss phenomenological aspects of a 
solution to the tachyonic slepton mass spectrum, based on contributions
 from D-terms to those soft SUSY breaking parameters.
We shall show that, depending on the specific  D-term charge assignments, 
the mass spectrum can be very different from the one obtained in the
 framework of a universal $m_0^2$ contribution to all the soft SUSY
 breaking squared mass parameters.

At the present stage, it is a trivial statement that the sum, 
\begin{equation}
\Sigma_{m^2_{AM}} \equiv (m_i^2)_{AM}+(m_j^2)_{AM}+(m_k^2)_{AM},
\end{equation}
is RG-invariant for  non vanishing Yukawa couplings,  $Y^{ijk} \neq 0$, 
because each of the soft scalar masses is RG-invariant.
Now, let us assume additional contributions to soft scalar masses 
$m_i^2$,
with the  requirement that the sum $\Sigma_{m^2_{AM}}$ does not change.
This can be realized by D-term contributions which are 
proportional to  charges $q_i$ of 
the chiral superfields under 
a broken symmetry, that is,  the total soft scalar 
mass is given by,
\begin{equation}
m_i^2 = (m_i^2)_{AM}+ q_im_D^2,
\label{Dterm}
\end{equation}
where $m_D$ is a universal parameter which defines the overall magnitude of 
the D-term contributions.
We require for the allowed Yukawa couplings that the total charge should be
conserved, i.e. $q_i+q_j+q_k=0$.
Hence, the sum does not change, 
\begin{equation}
M^2_{i,j,k} \equiv  m_i^2 + m_j^2 + m_k^2 = \Sigma_{m^2_{AM}},
\label{sum-0}
\end{equation}
and 
it is RG-invariant still after inclusion of the  additional D-term 
contributions. 
%
The only effect of the D term is to modify the boundary 
conditions of the scalar masses at the scale where the U(1)'s get broken, 
but with no effects on the RG evolution of these masses. 
Eq.  (\ref{Dterm}) is therefore valid at any scale.
For a fixed charge 
assignment, the  free parameters of the theory are $m_F$ and $m_D$.
Similarly, the sum, eq.(\ref{sum-0}), 
does not change if additional contributions 
to soft scalar masses are due to a certain type of supergravity theory,
e.g. moduli-dominated SUSY breaking in perturbative heterotic
 string models \cite{BIM,multi-T}.
In this case, the charge $q_i$ for the D-term contribution is replaced
by  $q_i\propto 1+n_i$, where  $n_i$ is the modular  weight of the
field $\Phi_i$. 
Hereafter, we mean charge $q_i$ as a coefficient of 
the deviation from the anomaly mediated soft SUSY breaking mass
parameters squared, with 
the universal
magnitude $m_D^2$ in eq.(\ref{Dterm}), to include the 
realization by the moduli-dominant SUSY breaking.
For the D-term contributions possible terms of $O(g^4)$ may also 
appear \cite{pomarol}.
Our assumption includes the fact that 
both the $(m_i^2)_{AM}$ and the D-term contribution, which is of 
$O(g^2)$, are the significant quantities and of comparable magnitudes.
Thus, we neglect further contributions of $O(g^4)$ compared 
with $(m_i^2)_{AM}$.
A similar statement has been done already in Ref.\cite{pomarol}.

\subsection{Discussion on models}
In order to proceed with our study, 
we need to assign the charges $q_i$.
For instance, in Ref. \cite{ibanez} it is clarified that 
there are three $U(1)$ symmetries, 
which are flavor-independent and allow the usual Yukawa couplings, 
i.e. $R$, $A$ and $L$ up to the baryon number symmetry $B$.
That is exactly consistent with the four degrees of freedom 
to deform each $m_i$ keeping the sum of the scalar squared masses of
the  superfields, 
$(Q,u,H_2)$, $(Q,d,H_1)$ and $(L,e,H_1)$ fixed.
The generic $U(1)$ symmetry $X$ is a linear combination of them,
\begin{equation}
X= mR + nA + pL + q B.
\end{equation}
We must then  identify the charge $q^X_i$ with $q_i$ 
for each chiral superfield $i= Q, u, d,
L, e, H_1$ and $H_2$. Charge assignments are shown in Table 1.
Observe that the hypercharge is just a linear combination of the 
$B$, $L$ and $R$ symmetries, $3 Y - 3 R - 3 L  = -B$. 

\begin{table}[htbp]
\begin{center}
\begin{tabular}{|c|c|c|c|c|c|c|c|}\hline
  & $Q$ & $u$ & $d$ & $L$ & $e$ & $H_1$ & $H_2$ \\ \hline
R &  0  & --1 &  1  &  0  &  1  &  --1  &  1    \\ \hline
A &  0  &  0  & --1 & --1 &  0  &   1   &  0     \\ \hline
L &  0  &  0  &  0  & --1 &  1  &   0   &  0    \\ \hline
B & --1 &  1  &  1  &  0  &  0  &   0   &  0     \\ \hline
$X$ & $-q$ & $-m+q$ & $m-n+q$ & $-n-p$ & $m+p$ 
& $-m+n$ & m \\ \hline
\end{tabular}
\vspace{.1in}
{\caption [0] 
{Charge assignment}.}
\end{center}
\end{table}

The extra $U(1)$'s appear naturally in GUT groups, like $SO(10)$
and $E_6$.
Indeed in the breaking $E_6\rightarrow SO(10)\times U(1)$ the extra
$U(1)$ 
charges have  the same sign for both left- and
right-handed lepton superfields\cite{hewett} providing a possibility to obtain 
positive slepton masses.
We will discuss this kind of model in detail in Section 3.1.

It turns out that constructing a viable model starting from $SO(10)$
is much more complicated.
One finds immediately that when $SO(10)$ breaks directly to the 
Standard Model gauge group or via $SU(5)\times U(1)$ with conventional
assignments of the fermions in five and ten dimensional
representations, one cannot get positive slepton masses squared.
On the other hand, when the breaking is via $SU(2)_L\times SU(2)_R\times
U(1)_{B-L}$, the charged slepton masses may become acceptable.

\subsection{Generic aspects}
Now let us discuss generic aspects of the sum rule.
For concreteness, we write explicitly the three types of  
sum rules, valid for all three generations,
\begin{eqnarray}
m_{\tilde Q}^2 +m_{\tilde u}^2+m_{H2}^2 &=& 
(m_{\tilde Q}^2 +m_{\tilde u}^2+m_{H2}^2)_{AM}, \\
\label{sum-m2-1}
m_{\tilde Q}^2 +m_{\tilde d}^2+m_{H1}^2 &=& 
(m_{\tilde Q}^2 +m_{\tilde d}^2+m_{H1}^2)_{AM}, \\
\label{sum-m2-2}
m_{\tilde L}^2 +m_{\tilde e}^2+m_{H1}^2 &=&
(m_{\tilde L}^2 +m_{\tilde e}^2+m_{H1}^2)_{AM}. 
\label{sum-m2-3}
\end{eqnarray}
Here we parameterize the deviations of the Higgs masses from 
the anomaly-mediated ones as
\begin{equation}
m_{Hi}^2=(m_{Hi}^2)_{AM}-d_{Hi}m_F^2.
\end{equation}

We fix the magnitudes of the $\mu$-term and the $B$-term  
by use of the minimization conditions for the Higgs effective
potential, to
assure  proper radiative electroweak symmetry breaking,  
\begin{eqnarray}
2\mu^2+M_Z^2  
 &=&{m_{H1}^2-m_{H2}^2 \over -\cos 2 \beta} -m_{H1}^2-m_{H2}^2, 
\label{mu}\\
2\mu B &=& \sin 2 \beta ({m_{H1}^2-m_{H2}^2 \over -\cos 2 \beta} -M_Z^2).
\end{eqnarray}

In the above, we have considered the minimization conditions 
derived from the tree level expression for the Higgs
effective potential.
The inclusion of the one-loop RG improved effective potential
would modify the above equations in such a way that the quantitative
behaviour of the solutions will be modified,
but the qualitative features are expected to be similar.
First of all, the condition 
$m_{\tilde L}^2+m_{\tilde e}^2 > 0$ requires 
$d_{H1} > 0.71$.
Another important condition for the successful electroweak symmetry 
breaking is the present experimental bound on $m_A^2$, 
where $m_A^2 = 2 \mu^2 + m_{H1}^2 +  m_{H2}^2 $ 
is the squared mass of the CP-odd Higgs  field.
For explicit models, which shall be discussed later, 
we require the fulfillment of the present experimental bound from LEP, 
$m_A > m_A^{exp} \simeq 88$ GeV \cite{expMH}. However, 
if $m_F^2$ is large enough compared with $M_Z^2$, 
the condition $m_A^2 >0$ is effectively equivalent to the experimental bound 
and corresponds to $\Delta m^2 \equiv
m_{H1}^2-m_{H2}^2 >0$ (see Eq. \ref{mu}).
In addition to the overall scale $m_F^2$,
the difference $\Delta m^2$ depends on $d_{H2}-d_{H1}$ and $\tan \beta$.
Thus, the condition  $\Delta m^2 >0$ 
leads to a minimum value of $d_{H2}-d_{H1}$, which depends on 
$\tan \beta$,  and 
combined with $d_{H1} > 0.71$ leads to a minimum value for $d_{H2}$.
In the absence of D-terms, $d_{Hi}=0$, $i=1,2$, we have 
$\Delta m^2/m_F^2 =O(10)$  except around  $\tan \beta \sim 50$.
Such a large value of $\Delta m^2$  implies that a negative value 
of $d_{H2}$ will be  allowed,
 even after including a non-vanishing value of  $d_{H1} > 0.71$.
On the other hand, around $\tan \beta = 50$ we have 
$\Delta m^2/m_F^2 =O(0.1)$ and then, for $d_{H1} > 0.71$, 
only  positive values of $d_{H2}$ 
are  allowed. The  smallness of $\Delta m^2$ close to $tan \beta = 50$
is due to the fact that for such large value of $\tan \beta$ the
bottom Yukawa coupling becomes strong and very close in  magnitude to the
top Yukawa coupling. Hence, the evolution of the Higgs mass parameters
is very similar,  $(m_{H1})_{AM} \simeq (m_{H2})_{AM}$, for 
$\tan \beta \simeq 50$. 
The solid line in Fig. 1 shows the minimum value of 
$d_{H2}$ against $\tan \beta$ under the condition 
$\Delta m^2 >0$, that is, the condition $m_A^2 > 0$ for 
$m_F^2 \gg M_Z^2$. For $\tan \beta > 50$ there is a change in the slope
for $d_{H2}$. This is  due to a change in the sign of the $\beta$-function
of the bottom Yukawa coupling after a vanishing value  is achieved 
due to a compensation  between the effects associated
with the strong gauge coupling  and the bottom
Yukawa coupling itself.

Now let us consider the minimum value of $m_F$.
The gaugino mass $M_2$ is obtained to be 
\begin{equation}
M_2 \simeq 0.43 \; m_F.
\label{eq:m2}
\end{equation}
As we shall show at the end of this section,
in the whole $(d_{H1},d_{H2})$ parameter space under consideration,  
we always have $M_2 \ll |\mu|$.
This implies that the lightest chargino is wino-like.
Hence, the present experimental
lower bound on the chargino mass, $m_{\chi^{\pm}} < 90 GeV$, implies that
the mass parameter $m_F$ is bounded to be 
\begin{equation}
m_F > 210~~{\rm GeV}.
\label{mF}
\end{equation}
For example, for $m_F=210$ GeV, the condition $m_A > m_A^{exp.} \simeq 88$ GeV 
lifts up the curve
which defines the minimum value of $d_{H_2}$ in Fig. 1 by 0.3.

Next we calculate the stop mass, in particular 
the average stop mass $m_{\tilde t,av}$, which is defined as 
$m_{\tilde t,av}^2 \equiv (m_{\tilde Q}^2+m_{\tilde u}^2)/2 + m_t^2$.
In the limit under discussion in this section, $m_F^2 \gg M_Z^2$,  
the contribution from the top quark mass is negligible and the  
average stop mass depends only on $d_{H2}$.
The dotted lines in Fig. 1 show contours of constant values of the ratio 
$R_{\tilde t}=m_{\tilde t,av}/M_2$ 
for $R_{\tilde t}=8, 10$ and 12.
Obviously, as  $d_{H2}$ increases, $R_{\tilde t}$ increases.
The thick solid line in Fig. 2 shows the minimum value of 
$R_{\tilde t}$ 
as a function of  $\tan \beta$.
The minimum value of $R_{\tilde t}$ is at  $\tan \beta  \simeq 3$ and
implies that $m_{\tilde t,av} \geq 6.6 M_2$. The thin solid line gives a
similar ratio for the average sbottom mass, 
$R_{\tilde b}=m_{\tilde  b,av}/M_2$ with 
$m_{\tilde b,av}^2 \simeq (m_{\tilde Q}^2+ m_{\tilde d}^2)/2 $.
The dotted line corresponds to a similar ratio but for first and second
generation squark masses  $R_{\tilde u}=m_{\tilde u,av}/M_2$ 
with $m_{\tilde u,av}^2 \equiv (m_{\tilde Q_{1,2}}^2 
+ m_{\tilde u_{1,2}}^2)/2$ being the average squared mass in the 
up-squark  sector. 
On can define the analogous 
quantity in the down sector
$R_{\tilde d}=m_{\tilde d,av}/M_2$, with $m_{\tilde d,av}^2 \equiv 
(m_{\tilde Q_{1,2}}^2 + m_{\tilde d_{1,2}}^2)/2$. 
It turns out, however, that the down-squark sector 
is stable as a function of $\tan \beta$,  with $R_{\tilde d}$ 
of order 10 for most
$tan \beta$  regions. Such a behaviour is expected since the main  
 dependence on $\tan \beta$ in the first and second generation 
down-squark  sector  
comes through   $d_{H1}$, which is fixed to its minimum 
value via the condition of positive slepton squared masses, $d_{H1} >
0.71$. The up-squark sector 
instead, depends on   $d_{H2}$ and hence on $\tan \beta$, as shown in
Fig 2. From  Fig. 2 it follows that,  all squark
masses are very heavy compared to gaugino mass parameters,
as expected from the underlying  structure of AMSB scenarios. 
This conclusion holds, unless there is a  large 
hierarchy   between the left and
right handed soft SUSY breaking parameters in the squark sector, as we
shall discuss below.

\begin{center}
\input fig-d2-tan.tex

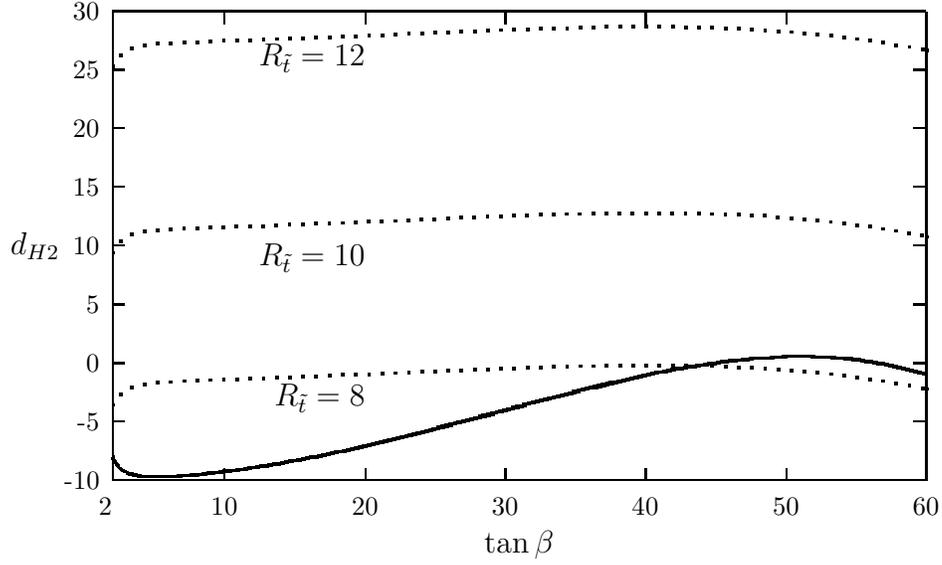
Fig.1: The minimum value of $d_{H2}$ (solid line) and constant contours 
of $R_{\tilde t}$,  the ratio of the
average stop mass to the SU(2) gaugino mass parameter, $M_2$, which
determines the chargino mass in this large $\mu$ scenario,
(dotted lines).
\end{center}

\begin{center}
\input fig-squ-tb-min.tex

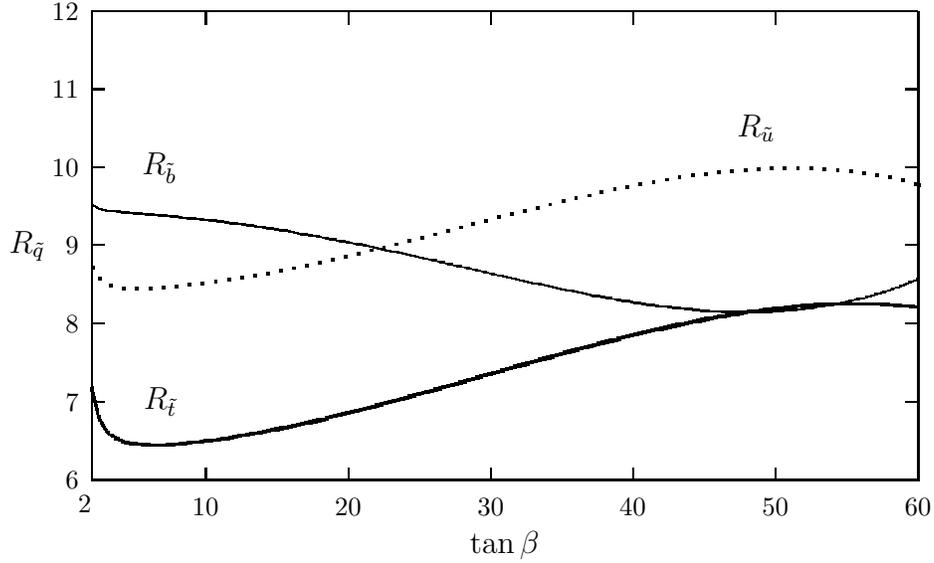
Fig.2: The minimum of the ratio of the average stop mass 
(thick solid line),  of the 
average sbottom mass  (thin solid line) and 
of the first and second generation up-sector squark masses
(dotted line) to the SU(2) mass parameter $M_2$.
\end{center}

\vskip 1cm
Now let us discuss the magnitude of $\mu$ in these models.
In the case with $d_{Hi}=0$, we have 
\begin{equation}
{\mu^2+M_Z^2/2 \over m_F^2} \sim 10,
\label{mu-2}
\end{equation}
 for any value of $2 \leq \tan \beta \leq 60$.
 Moreover, from the expression for $\mu^2$  
\begin{equation}
\mu^2=-m_{H1}^2+[(m_A^2+M_Z^2){\tan ^2 \beta \over \tan ^2\beta +1} -
M_Z^2],
\end{equation}
in which  the second term in RHS is always positive for 
$\tan \beta >M_Z/m_A$,
one can derive the following inequality,
\begin{equation}
\mu^2 > -m_{H1}^2 > -(m_{H1}^2)_{AM}.
\label{mu-3}
\end{equation}
The second inequality in eq.(\ref{mu-3}) is due to the constraint 
$d_{H1} >0$.
We combine eqs. (\ref{eq:m2}), (\ref{mu-2}) and (\ref{mu-3}), and 
find that $|\mu|$ is larger than $M_2$.
Fig. 3 shows the minimum value of the ratio 
$R_\mu \equiv \mu'/M_2$, with $\mu'=\sqrt{\mu^2 +M_Z^2/2}$, as a
function of $\tan\beta$,  
in the parameter space $(d_{H1},d_{H2})$ allowed 
by the conditions $m_{\tilde L}^2+m_{\tilde e}^2 > 0$ 
and $\Delta m^2 >0$.
To realize the minimum value, an extreme value of the ratio 
$r_d=d_{H2}/d_{H1}$ is sometimes required.
For example, for $\tan \beta =3$ the minimum value of $R_\mu \simeq 2.5$ is 
obtained, and that is realized for $r_d=-13$.
Fig. 4 shows the minimum value $R_\mu$ as a function of  $r_d$ for 
$\tan \beta =3, 20$ and 50.
Note that as $\tan \beta$ increases, the minimum values of $d_{H2}$ 
and $r_d$ increase.
Hence, as a generic aspect  of these models, the mass parameter
$|\mu|$ (as well as the  the squark masses) is large compared with 
the gaugino mass parameters, $M_1$ and $M_2$.

\begin{center}
\input fig-min-mu-tan.tex

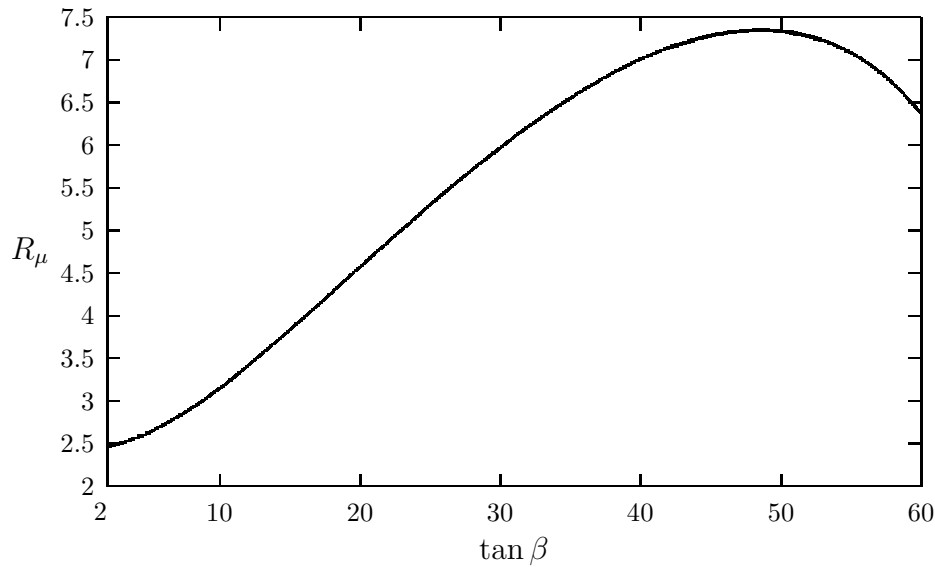
Fig.3: The minimum value of $R_\mu = \sqrt{\mu^2 +M_Z^2/2}/M_2$ as a
function of $\tan \beta$.
\end{center}

\begin{center}
\input fig-min-mu-3-50.tex

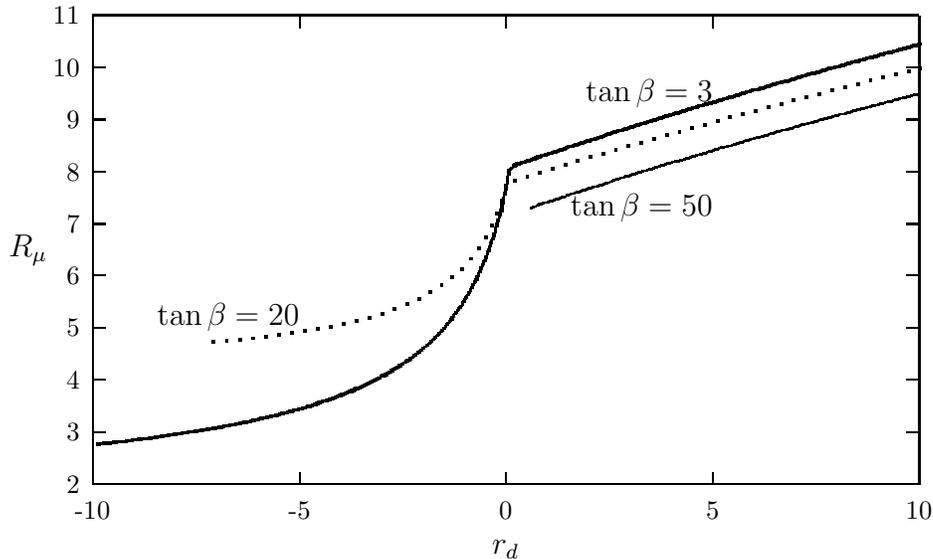
Fig.4: The minimum of the ratio $R_\mu$ as a function of the ratio 
$r_d = d_{H2}/d_{H1}$.
\end{center}

\section{Mass Spectrum}

In order to show in detail possible  mass spectra in generalized AMSB
scenarios with extra D-term contributions to the sfermion mass
parameters,  we discuss two explicit  examples in this section.

\subsection{A simple example}

First we consider a simple case, where 
additional contributions are degenerate for the sfermions and for the
Higgs mass parameters, respectively,
\begin{equation}
m_{\tilde f}^2 = (m_{\tilde f}^2 )_{AM} +m_D^2, 
\qquad 
m_{Hi}^2 = (m_{Hi}^2 )_{AM} -2m_D^2.
\end{equation}
This type of charge assignment can be realized through 
the breaking $E_6 \to SO(10) \times U(1)$.

In the following we shall calculate $m_A$.
Note that in this example $m_D$ does not contribute to $\Delta m^2$ 
or $m_A$ and hence
the mass  $m_A$ is determined by $m_F$ and $\tan \beta$.
The result is shown in Fig. 5, where the region above the solid line
defines the condition  $m_A > 88$ GeV.
As a result, $m_F > 270$ GeV is required around $\tan \beta =50$.
In most of the $\tan \beta$ region, the requirement $m_A > 88$ GeV is less 
significant than  the experimental lower 
bound on the chargino mass, $m_F > 210$ GeV.
The dotted lines in Fig. 5 
correspond to $m_A=250, 500, 750$ and 1000 GeV, respectively.

\begin{center}
\input fig-mA.tex

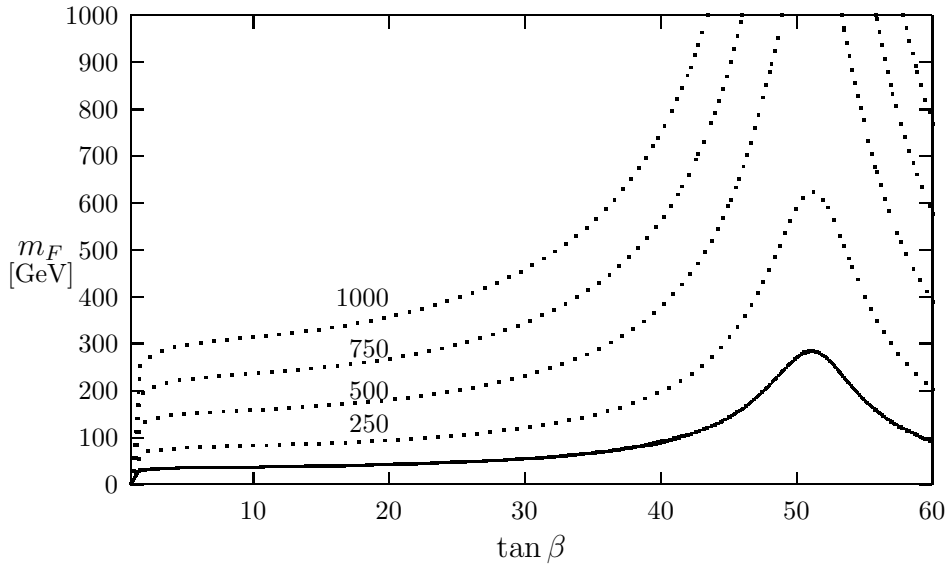
Fig. 5: Lines of constant values of the CP-odd Higgs mass, $m_A= 88,
250, 500, 750$ and $1000$ GeV, in the $m_F$--$\tan \beta$ plane.
\end{center}

Now we discuss the slepton masses and the Higgsino mass parameter $\mu$,
in both the small and large $\tan \beta$ scenarios.
For example, we take $\tan \beta =3$ and 50.
Fig. 6 shows $\mu$ and $m_{\tilde L}$ for $\tan \beta =3$.
In this case  the masses 
$m_{\tilde L}$ and $m_{\tilde e}$ 
are almost degenerate for all three generations, since   the
corresponding Yukawa couplings contributing to Eq. \ref{scalar-m} are very
small. 
The dotted lines correspond to $|\mu|=0.2, 0.5, 1.0, 1.5$ 
and 2.0 TeV.
The solid lines correspond to 
$m_{\tilde L}=m_{\tilde e}= 85, 100, 200, 400$ and 600 GeV.
The region below  $m_{\tilde L}=m_{\tilde e}= $ 85
GeV is excluded by present bounds on the smuon  mass
from the combined results of the four LEP 
experiments \cite{lepsusy}\footnote{ The
  analogous lower bound on the selectron mass is about 90 GeV. The
 limits quoted here for slepton masses include data up to $\sqrt{s} = 189$
GeV. The inclusion of the latest data up to  $\sqrt{s} = 202$ GeV
will improve the bounds on smuons and selectrons by about 5 GeV, 
respectively.  The
precise value of the experimental bound is not crucial for the general
analyses performed  in this paper.}.
The region to the left of the dot-solid line is excluded due to the
present experimental bounds on the chargino mass.

\begin{center}
\input fig-lmu3.tex


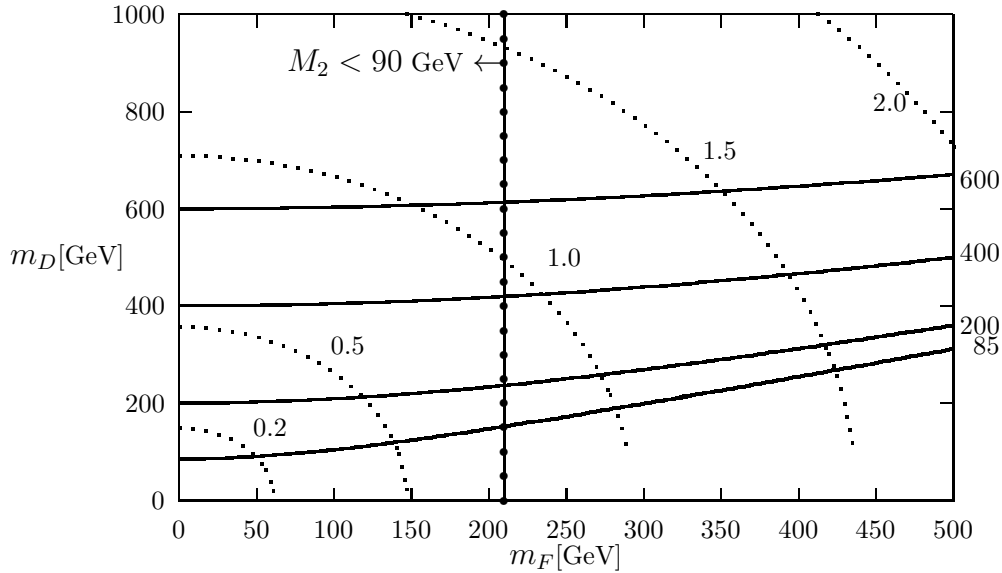
Fig.6: Contours of constant values for the smuon masses for  
$m_{\tilde L}=m_{\tilde e}= 85, 200, 400, 600$ GeV  (solid
lines)  and for   $\mu = 0.2, 0.5, 1, 1.5$ TeV
(dotted lines) in the $m_F$ - $m_D$ plane, for $\tan \beta =3$. 
\end{center}

The first and second families of slepton masses for $\tan \beta =50$ 
are almost the same as those for $\tan \beta =3$. This is the case since,
although  the off-diagonal elements of the slepton mass matrices  have 
a $\tan \beta$ enhancement factor, they are also
proportional to the first and second
generation lepton masses which are too small. 
Hence, ignoring the small $SU(2)_L \times U(1)_Y$ D-term
contributions,
 the two mass eigenstates, for the selectrons and smuons
are still approximately given by 
$m_{\tilde L}$ and $m_{\tilde e}$, respectively, independent of $\tan \beta$.

On the other hand, for large $\tan \beta$ the stau mass matrix has a 
sizable off-diagonal element, 
which reduces the eigenvalue of the lightest stau mass.
For the same values of the parameters $m_F$ and $m_D$, the stau mass, 
$m_{\tilde \tau_1}$, is  smaller than 
the first and second generation  slepton masses and hence it
 excludes a wider region of that parameter space. In the following we
consider 
 the experimental lower bound on the lightest stau mass from LEP,
$m_{\tilde \tau_1} > 70 GeV $ \cite{lepsusy}.
Fig. 7 shows the curves corresponding to the  lightest
stau mass,  
$m_{\tilde \tau_1}=70, 200, 400$ and 600 GeV, as well as 
the curves for constant values of $|\mu|$ and the curve corresponding
to the experimental lower bound,  $m_{\tilde L}=m_{\tilde e}= 85$ GeV,
 for the smuons as a reference. 
The region to the left of the dot-solid line corresponds to values of the
  CP-odd mass $m_A$ which are experimentally excluded by LEP.
The variation of the lightest stau  mass, $m_{\tilde \tau_1}$ with the
sign of  $\mu$ is negligible. In Fig. 7 the value of $|\mu|$ as a 
function of $m_F$  is slightly  smaller than in the 
low $\tan \beta$ case of Fig.6.

\begin{center}
\input fig-stau50.tex

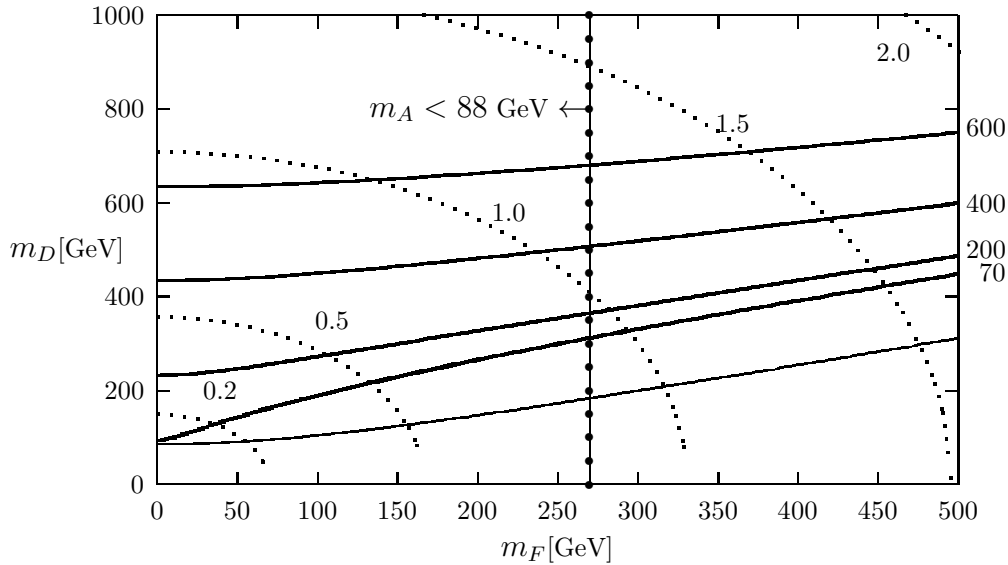
Fig.7: Contours of constant values for the stau mass,
  $m_{\tilde  \tau_1} = 70, 200,400, 600$ GeV for $\tan \beta
=50$  (thick solid lines) and for  $|\mu|= 0.2, 0.5, 1.0, 2.0$ TeV
(dotted lines). The thin solid line shows the lower experimental bound 
on the smuon mass, $m_{\tilde L} \simeq m_{\tilde e} \simeq $ 85
  GeV.
\end{center}

Similarly we can discuss the predictions of this model for the stop sector.
For the stop mixing angle $\theta_t$, $\sin 2\theta_t$ is always large.
That is because $m_{\tilde Q}^2-m_{\tilde u}^2$ for the stop 
is small compared 
with $|\mu|$ as well as $|A_t|$.
For example, we have $\sin 2\theta_t > 0.8$ for $\tan \beta =3$, 
$m_F < 500$ GeV and $\mu >0$.
The negative sign of $\mu$ leads to a slightly larger value 
of $\sin 2\theta_t$.
For $\tan \beta = 50$ we have $\sin 2\theta_t=1$, with 
$m_F \leq O(1)$ TeV.
Figs. 8 and 9 show the lightest stop mass for $\tan \beta =3$ and 50 
for  $\mu <0$.
The solid lines correspond to different values for the lightest stop
mass,  $m_{\tilde t_1}= 0.09, 0.2, 0.5, 1.0$ and 
1.5 TeV.
Present experimental bounds exclude the region  
$m_{\tilde t_1} < 90 $ GeV \cite{lepsusy}.
In Fig. 9 ($\tan \beta =50$), the excluded experimental bound on the
lighetst stop mass is not shown,
because  it corresponds to a very narrow region already excluded by the
stau mass constraint.
The experimental bound of 
the chargino mass excludes the area to the left of the dot-solid curve
and gives a stronger constraint, Eq. (\ref{mF}), which 
implies that 
the minimum of the $m_{\tilde t_1}$ is about 500 GeV for 
$\tan \beta =3$.
Furthermore, for $\tan \beta =50$ we have the constraint due to 
$m_A$, 
i.e. $m_F > 270$ GeV, which excludes the region to the left of
the dot-solid line in Fig. 9.
Thus,
in the case of large $\tan \beta$, with similar values of the bottom and
top Yukawa couplings,  the minimum value of $m_{\tilde t_1}$ is about 800 GeV. 
The case with $\mu >0$ leads to a slightly larger mass 
of the lightest stop.

In the  case where a universal mass $m_0$ is added to solve the
problem of tachyonic solutions in the slepton sector, the particle
spectrum is such that 
the lightest stop is quite heavy and $|\mu|$ is large. 
Hence, the present example is phenomenologically not very different 
from the model with the universal $m_0$ addition.
This simple  example shows  two features which are quite generic in this type
of generalized AMSB models.
One is that the lightest stop is in general quite heavy and 
the other is that for most $\tan \beta$ values (except around 
$\tan \beta =50$), the chargino is predicted to be so light that
present experimental bounds on chargino masses put constraints on the
allowed values of  $m_F$, which are stronger than those derived from
demanding a successful electroweak symmetry breaking, 
with  $m_A> 88$ GeV.

\begin{center}
\input fig-stop-.tex

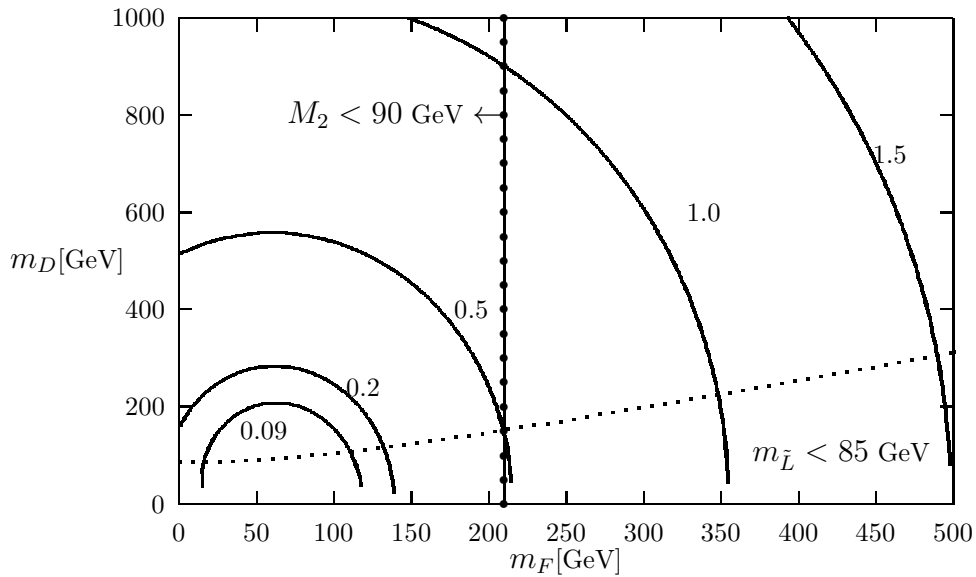
Fig. 8: Contours of constant   lightest stop mass, $m_{\tilde t_1} =
   0.09, 0.2, 1, 1.5$ TeV, in the $m_D$--$m_F$ plane,
for $\tan \beta =3$ and $\mu <0$. The regions 
below the dotted line and
to the left of the dot-solid line are experimentally excluded.
\end{center}

\begin{center}
\input fig-stop-50.tex

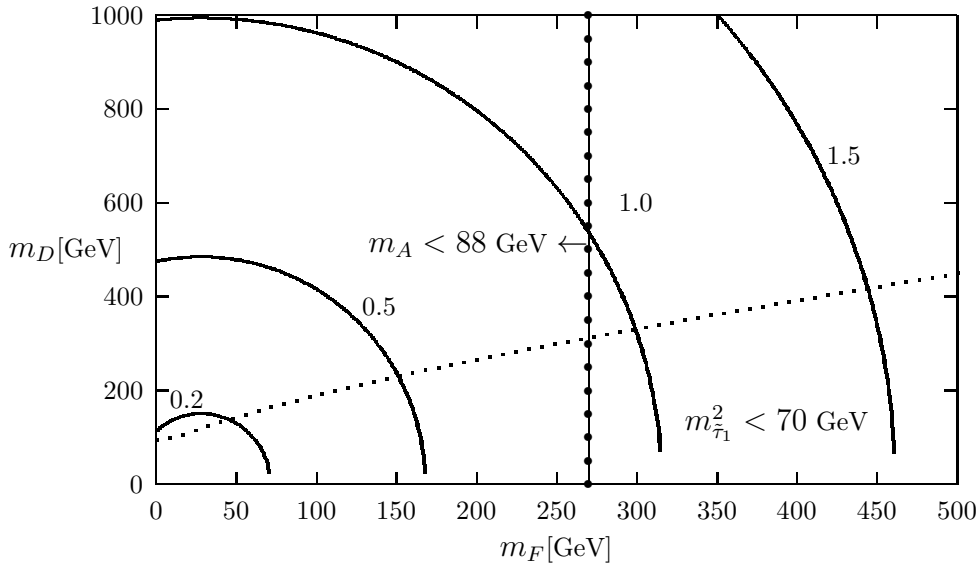
Fig. 9: Contours of constant   lightest stop mass, $m_{\tilde t_1} =
   0.2, 1, 1.5$ TeV, in the $m_D$--$m_F$ plane,
 for $\tan \beta =50$ and  $\mu <0$
The same as in Fig. 8, the regions
below the dotted line and
to the left of the dot-solid line are experimentally excluded.
\end{center}

It is interesting to notice that, in most of the allowed regions of 
parameters, the 
ligthest supersymmetric particle (LSP) is the wino-like neutralino 
and the next-to-LSP (NLSP) is the chargino.
We have here the same spectrum for charginos as
in minimal AMSB scenarios: LSP wino-like and almost 
degenerate with the chargino making detection 
difficult \cite{anom-ph,gunion,cheng}.
In addition, in the region close to the region where the lightest stau mass  
is close to its experimental limit, 
we may have the stau to be the  LSP, although 
such region is narrow against $m_{D}$.
A charged, stable LSP is cosmologically disfavoured. However, we can
avoid this problem by demanding $m_{\tilde \tau_1} > M_2$ and none 
of the features discussed in this section  will vary in any
significant manner.

\subsection{A model with a light stop }

Here we consider a special case in which, by  significantly shifting the
degeneracy between the left and right handed stop/sbottom soft SUSY
breaking parameters we obtain a light stop in the spectrum.
%
Note that, in general, 
the supersymmetric spectrum is constrained by direct experimental
searches and by the requirement that  it provides a good description
of the precision electroweak data. This requirement implies that,
to avoid an unacceptable large contribution from supersymmetric
particle loops to the $\rho$ parameter, and 
unless unnatural cancellations take place, the soft SUSY
breaking mass parameters for the left-handed top squark should be
larger that 300 GeV \cite{CCRW}. 
Quite generally, the heavier the supersymmetric 
spectrum, and in particular the heavier the left-handed sfermions, the better
the agreement between the MSSM and the precision electroweak observables. 
Hence, in the following, we shall consider a case with a light stop,
which is mainly right-handed, so that the model is not in any conflict
with precision measurements.

First, let us discuss the charge assignment leading to a light stop.
We take the sign assignment of charges,  $q^X_i \equiv X_i$, 
such that $m_D^2 >0$ in eq.(\ref{Dterm}).
Obviously, the minimal requirement that  
$m_{\tilde L}^2 > 0 $ and $m_{\tilde e}^2 > 0$ 
implies  $X_L=-n-p > 0$ and $X_e=m+p > 0$, which also yields 
 the condition $X_{H1} = -m+n < 0$.
In order to achieve proper electroweak symmetry breaking, a large 
$\Delta m^2$ is desirable, and then we need 
$X_{H1}-X_{H2} = -2m+n \geq 0$.
That also implies that $X_{H2} = m < 0$.
Combining all the previous conditions we have 
\begin{equation}
m > -p > n \geq 2m.
\end{equation}
The inequality $X_{H2} <0$ requires $X_Q+X_u =-m> 0$.
In order to obtain a light stop, 
either  $X_Q$ or $X_u$ should be negative.
Therefore, if $Q$ and $u$ belong to a multiplet of 
a larger gauge group like $SU(5)$, ($X_Q = X_u$),
the present scenario predicts heavy stops.
If $X_Q <0$, then the lightest stop would be left-handed and most
probably in conflict with present constraints from precision
measurements, 
unless its mass is sufficiently close to that of the left-handed
lightest  sbottom 
which would also  appear in the spectrum.
Alternatively, we have the possibility $X_Q > 0$ and $X_u <0 $, which may
lead  to a right-handed lightest stop.
In any case, most sign assignment of charges are fixed in a light 
stop model.

As an example we take $(m,n,p,q)=(-2,-4,3,-3)$, i.e., 
\begin{equation}
(X_Q,X_u,X_d,X_L,X_e,X_{H1},X_{H2}) = 
(3,-1,-1,1,1,-2,-2).
\end{equation}
In this case the CP-odd mass $m_A$ is independent of $m_D$ and the 
behaviour of the slepton masses
 is similar to the case in which  all the squarks are quite heavy.
Figs. 10 and 11 show the lightest stop mass
for $\tan \beta =3$ and 50 and $\mu < 0$. For the small $\tan\beta$
case, there is an allowed region in  
the $m_D$--$m_F$ plane  where, even after 
imposing the chargino mass bounds on  
$m_F$, $m_F \geq 210$ GeV,
there are solutions  which allow for a very light stop. 
In the case $\tan \beta \simeq 50$,
for which $m_{H1}^2 \simeq m_{H2}^2$  and  a more stringent bound on
$m_F$ follows from the experimental bound on $m_A$,
only heavy stops are allowed unless $m_D \geq 1$ TeV.
The solid lines in Fig. 10 and 11 correspond to 
$m_{\tilde t_1}= 0.09 , 0.2, 0.5, 1.0$  and 1.5 TeV.
The region to the left of  the curve $m_{\tilde t_1}= 90$ GeV 
corresponds to 
the experimentally excluded region of stop masses which excludes an
important region of parameter space even for very large values of  $m_D$.
The region below the dotted line,  
$m_{\tilde L} < 85$ GeV and  $m_{\tilde \tau} < 70$ GeV,
Figs. 10 and 11 respectively,  is experimentally excluded.
For  $\tan\beta \simeq 50 $ and fixed values of $m_D$ and $m_F$,
the lightest sbottom has a mass of similar magnitude  to the lightest 
stop mass shown in Fig. 10.

\begin{center}
\input fig-stop-u1-.tex

Fig. 10:  Contours of  the lightest stop mass, $m_{\tilde t_1}= 0.09 ,
0.2,  0.5,  1.0$ TeV, (solid lines) in the
$m_D$--$m_F$ plane, for 
$\mu <0$ and   $\tan \beta =3$. 
The regions below the dotted line, to the
left of the dot-solid line and to the left of the line of $m_{\tilde
  t_1}= 90$ GeV are experimentally excluded. 
\end{center}

\begin{center}
\input fig-stop-u1-50.tex

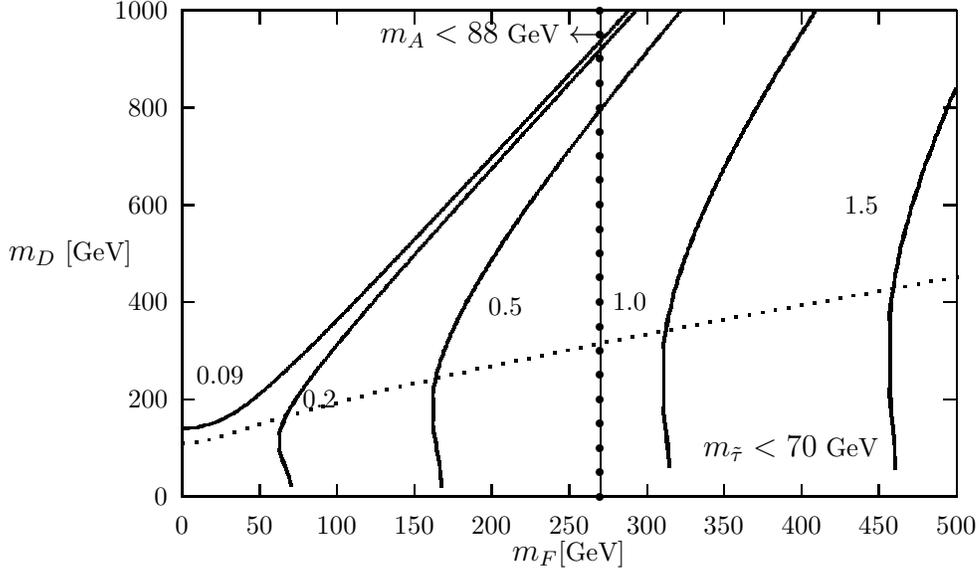
Fig. 11:  Contours of  the lightest stop mass, $m_{\tilde t_1}= 0.09 ,
0.2, 0.5, 1.0, 1.5$ TeV (solid lines) in the
$m_D$--$m_F$ plane, for 
$\mu <0$ and   $\tan \beta =50 $. 
Analogous to Fig 10, the regions below the dotted line, 
to the left of the dot-solid line  and  to the left  of the line of 
$m_{\tilde t_1}= 90$ GeV 
are experimentally excluded. 
\end{center}

Similarly, we can discuss other cases leading to a light 
$\tilde t_1$.
For example we can vary $q$ fixing $(m,n,p)=(-2, -4, 3)$ as in the
previous case.
Fig. 12 shows the lightest stop mass values as a function 
of $m_D$ for $m_F=250 $ GeV  and $\tan \beta$ = 3, for 
$q=-3, -4$ and $-5$.

\begin{center}
\input fig-stop-d.tex

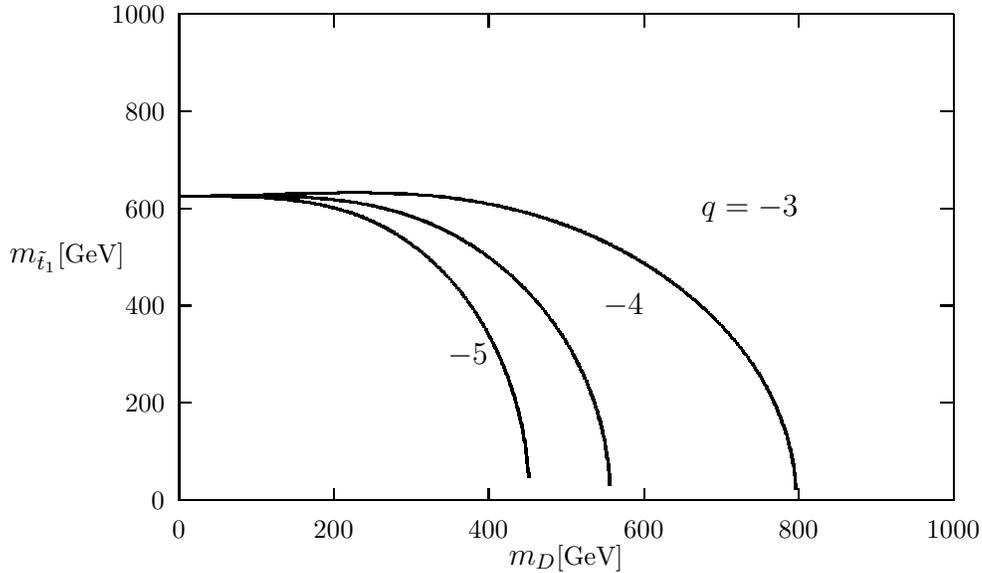
Fig. 12 The lightest stop mass as a function of  $m_D$, for $m_F$ =
250 GeV, $\mu <0$ and $\tan \beta =3$ with $q=-3, -4$ and $-5$, respectively. 
\end{center}

\section{Conclusions}

We have studied phenomenological aspects of a generalization of 
anomaly-mediated 
SUSY breaking scenarios with non-decoupling effects from a high scale
theory, which allow for D-term contributions to be effective at the
low energy scale. 
We have assumed that the additional D-term contributions to 
the soft supersymmetry breaking scalar mass parameters, $m_i$, are
such that  
the sum $m_i^2+m_j^2+m_k^2$, corresponding to chiral fields in 
the allowed Yukawa couplings, $Y^{ijk} \neq 0$, 
remains  RG-invariant as it occurs in the AMSB
models.
The extra D-term contributions, depending on the charge assignment of
the extra U(1)'s involved in the model, can solve the problem of
tachyonic solutions in the slepton sector, whereas preserving the
flavour independence of the solutions.
Most interesting, since the RG-invariant sums of non vanishing
Yukawa couplings  $Y^{ijk} \neq 0$ appear directly in   
the renormalization group evolution of the Yukawa couplings and
soft SUSY breaking masses themselves, one could allow for additional
contributions to   each of the soft scalar masses and,  as long as 
the sum itself is not  affected, 
the soft scalar masses will
remain RG-invariant. Given a fixed charge assignment, the sparticle
spectrum is uniquely determined as a function of 
the  free parameters $m_F$ and $m_D$, $\tan \beta$ and the sign of
$\mu$.
In general, the average stop mass and the parameter $|\mu|$ are 
larger  than the slepton 
masses and the gaugino mass parameters $M_1$ and $M_2$, as expected 
from the underlying AMSB structure.
The  mass spectrum in the most simple case is similar to the case 
in which a universal  contribution $m_0^2$ is added, to all the soft SUSY
breaking scalar mass parameters squared, to cure the
tachyonic mass problem.
However, as we have explicitly shown, it is possible to construct
models in which a light stop, compatible with present electroweak
precision measurements  will naturally appear in the spectrum.
Models with light third generation squarks demand  a specific  U(1) charge 
assignment and yield constraints for the
model building.

\vspace{1cm}

{\bf Note added:}
After completion of this work, an article \cite{jj2} appeared, 
where $D$-term contributions and sum rules of soft scalar masses 
are also discussed.

\vspace{1cm}

{\bf Acknowledgements}:
The work of KH is partially supported by the Academy of Finland
project no. 163394.

\end{document}